\documentclass{article}
\usepackage[final]{nips_2017}

\usepackage[utf8]{inputenc} 
\usepackage[T1]{fontenc}    
\usepackage{hyperref}       
\usepackage{url}            
\usepackage{booktabs}       
\usepackage{amsfonts}       
\usepackage{nicefrac}       
\usepackage{microtype}      

\usepackage{array,amsmath}
\usepackage{algorithm}
\usepackage{algorithmic}
\usepackage{graphicx}

\title{Sequential Matrix Completion}

\author{
  Anne Marsden 
     \\
  Department of Computer Science\\
  University of Chicago\\
  Chicago, IL, 60637 \\
  \texttt{marsden@stanford.edu} \\
   \And
   Sergio Bacallado \\
   Statistical Laboratory \\
   University of Cambridge \\
   Cambridge, UK, CB30WB \\
  \texttt{sb2116@cam.ac.uk} \\
}

\begin{document}

\maketitle

\begin{abstract} 
We propose a novel algorithm for sequential matrix completion in a recommender system setting, where the $(i,j)$th entry of the matrix corresponds to a user $i$'s rating of product $j$. The objective of the algorithm is to provide a sequential policy for user-product pair recommendation which will yield the highest possible ratings after a finite time horizon. The algorithm uses a Gamma process factor model with two posterior-focused bandit policies, Thompson Sampling and Information-Directed Sampling. While Thompson Sampling shows competitive performance in simulations, state-of-the-art performance is obtained from Information-Directed Sampling, which makes its recommendations based off a ratio between the expected reward and a measure of information gain. To our knowledge, this is the first implementation of Information Directed Sampling on large real datasets. 
 This approach contributes to a recent line of research on bandit approaches to collaborative filtering including \citet{rctest}, \citet{li2010contextual}, \citet{bresler2014latent}, \citet{li2016collaborative}, \citet{deshpande2012linear}, and \citet{zhao2013interactive}. The setting of this paper, as has been noted in \citet{rctest} and \citet{zhao2013interactive}, presents significant challenges to bounding regret after finite horizons. We discuss these challenges in relation to simpler models for bandits with side information, such as linear or gaussian process bandits, and hope the experiments presented here motivate further research toward theoretical guarantees. 
\end{abstract} 

\section{Introduction}
\label{intro}
A recommender system or collaborative filter uses a database of user preferences to make sequential product recommendations. Let $M \in \mathbb{R}^{D\times N}$ be a matrix containing the true user preferences, where each row corresponds to a user and each column corresponds to a product. Our database consists of a sequence of entries of $M$ observed with noise
\begin{equation}
Y_{t}=\textrm{tr}(Z_{t}^{\top}M) + \varepsilon_{t},
\end{equation}
where $Z_{t} \in \left \{ e_{i}e_{j}^{\top} ; i \in [D], j \in [N] \right \} $, and $\varepsilon_{t}$ is the noise in the observation. We denote the set of entries observed $\mathcal{Z}=\left \{ Z_{t}  \right \}$.

Much research on recommender systems has focused on the matrix completion problem. A body of work initiated by \citet{Candes-Recht} has shown that, if the matrix $M$ is of low-rank---user preferences are explained by a few latent features---and the factors of the matrix satisfy certain decoherence conditions, it is possible to complete the matrix from a small set of possibly noisy observations. Our aim here is to study the matrix completion problem sequentially. Briefly, we would like to find a sequential rule or policy for product recommendation which  will yield the highest possible ratings after a finite time horizon. 

More formally, suppose at each step we can choose an action $Z_{t}$ that corresponds to observing some, possibly corrupt, entry of the matrix.  
Here, the source of corruption would be that the user does not rate the product consistently. We define the reward 
for taking action $Z_{t}$ at step t as
\begin{equation}
r_{t,Z_{t}}=(\textrm{tr}(Z_{t}^{\top}M)+\varepsilon_{t})\beta^{T_{t,Z_{t}}},
\end{equation}
where $(\varepsilon_{t})_{t\geq1}$ is a sequence of independent noise variables with mean 0 and variance $\sigma^{2}$. The variable $T_{t,Z_{t}}$ counts the number of times that action $Z_{t}$ has been chosen before time $t$, and the parameter $\beta$ geometrically discounts the reward of an entry after it has been observed. 
We will focus on the case $\beta=0$, which for $M$ non-negative corresponds to the case where we can make a certain recommendation at most once.

At each step $t$, the experimenter observes the reward $r_{t,A_{t}}$ for the action chosen, $A_{t}$.
We define the pseudo-regret for finite-horizon $T$ as
\begin{equation}
R_{T}(A)= \sup \limits_{Z_{1}, \dots, Z_{T}} \mathbb{E} \bigg( \sum  \limits_{t=1}^{T} r_{t,Z_{t}} \bigg)- \mathbb{E} \bigg( \sum \limits_{t=1}^{T} r_{t,A_{t}} \bigg) .
\end{equation}
The goal of our recommender system will be to minimize the pseudo-regret, especially when $T$ corresponds to a number of entries smaller than the matrix completion threshold---the horizon at which the matrix can be completed confidently with a semidefinite program. 

The problem described is related to contextual bandits, a family of multi-armed bandits including linear and gaussian process bandits in which the mean reward of an arm is a function of a set of given predictors. The principal difference is that in sequential matrix completion, we must learn latent factors which explain the rewards. The formulation of the problem above is meant to emphasize this distinction. However, it would be straightforward to adapt the algorithms introduced in this paper to the requirements of realistic recommender systems, such as (i) taking advantage of predictors associated to users or products, (ii) restrictions of the set of user-product pairs available at any given step, (iii) models for non-response, and (iv) prior information about the matrix of preferences.

	As a note with regard to this formulation, in certain applications, it might not be possible to choose the user at each step. At an abstract level, our algorithm performs Bayesian optimisation over sets indexed by two categorical variables, which has applications beyond recommender systems. Furthermore, this paper will later show that the ability to choose user-item pairs can be exploited for information gain that achieves remarkable performance. In real recommender system settings, our formulation  makes it convenient to include information on the likelihood that a user gives any feedback into the model— this helps ease the restrictive assumption that a user will definitely respond with their rating of the item. It can thus highlight whether there exist certain users that are good to interact with to gain more information. However, in this formulation one must be careful to avoid the pitfall of recommending items only to users who tend to give higher ratings. To avoid this, one could adaptively scale the user-columns so as to equalize their average rating. Finally, note that if there were constraints on the set of users or items available for recommendation at each time step, these could be imposed naturally with our formulation. 

The remainder of the introduction provides some background on stochastic bandit policies. Section 2 defines two policies for sequential matrix completion based on a gamma process factor model, and describes a fast variational procedure for inference. Section 3 evaluates the policies in simulations using synthetic data, as well as 3 real datasets. The final section provides further connections to the literature and discusses theoretical challenges.

\subsection{Exploration-Exploitation Trade-offs and Stochastic Bandit Policies}
Our estimator policy relies on an accurate reconstruction of the complete matrix, $M$, knowing only observations corresponding to $\mathcal{Z}$. Note that this is only feasible when $M$ is low rank. Matrix completion algorithms
typically need the observed entries to be distributed somewhat uniformly around the matrix. For instance, an estimator policy that only interacts with one user would have no hope of completing the matrix, as it knows nothing about the other users' preferences. In this sense the estimator policy must explore the user-item pairs. On the other hand, the policy must begin to exploit user-item pairs that are expected to be optimal in order to minimise the regret.

This type of problem relates to well-studied strategies on the multi-armed bandit problem.  The narrative for this problem is as follows.  A player is at a casino with multiple slot machines. 
 At each step he chooses an arm to pull and then observes the reward of his choice, which is assumed to be drawn from some underlying distribution associated to the chosen arm. To perform well the player must strike a balance between exploration and exploitation. That is, he must pull each arm enough to get a rough estimate of their expected reward, but he must also exploit the arms that seem to have high rewards. 
 
Suppose there are $K$ arms with underlying parameters $\theta_{1}, \dots, \theta_{K} \in \Theta$ that respectively describe the distribution of rewards for each arm. At time $t$ we choose an arm $A_{t}\in[K]$ which depends on the history of rewards before $t$ according to some policy. The reward $X_{t}$ is drawn from the distribution $\nu_{\theta_{A_t}}$ with expectation $\mu(\theta_{A_t})$. The measure of performance for a policy is the cumulated regret, which at time $t=n$ is
 \begin{equation}
 \label{freq:regret}
 R_{n}(\theta)= 
 \mathbb{E}_{\theta} \Big[ \sum \limits_{t=1}^{n} \mu^{*} - X_{t}  \Big] = 
 \sum \limits_{t=1}^{n} \mu^{*} - \mu(\theta_{A_t}) ,
 \end{equation}
 where $\mu^{*}= \textrm{max} \left \{ \mu(\theta_j), j \in [K] \right \}$. Frequentist analyses of multi-armed bandits have focused on worst-case bounds---upper bounds on $\sup_{\theta\in\Theta} R_n(\theta)$ for specific policies---as well as minimax results, which lower bound this quantity over a space of policies. Bayesian analysis of regret, on the other hand, assumes the parameter $\theta$ is random and bounds the expected regret under a prior distribution $\pi$,
\begin{equation}
R_{n}= \mathbb{E}_{\pi} \Big[ R_{n}(\theta) \Big],
\end{equation}
also known as the Bayes regret. The policy which minimises Bayes regret is known as Bayes-optimal and is typically the solution to an intractable dynamic program. An important exception is the case of a geometrically discounted multi-armed bandit, in which the Gittins index strategy is provably Bayes-optimal \citep{weber1992gittins}.

An effective and simple strategy from both Bayesian and frequentist perspectives is Thompson Sampling (TS). Let $ \pi_{t}$ denote the posterior distribution after $t-1$ observations. We draw $(\theta_{i,t})_{i\in[K]} \sim \pi_{t}$, and then choose an action $A_{t} \in \textrm{arg max}_i\, \mu(\theta_{i,t})$. In other words, an arm is chosen with a probability equal to the posterior probability that it is the best arm. The worst-case and Bayes regret of TS have been characterised in a range of models, and it is known that in the case of Bernoulli rewards, the regret grows at the optimal rate \citep{agrawal2013further, kaufmann2012thompson, russo2014learning}.

Policies like TS or the Upper Confidence Bound method \citep{lai1985asymptotically} encourage exploration through the heuristic of optimism under uncertainty, but they do not explicitly quantify the information to be gained by each possible actions. 
By contrast, Information Directed Sampling (IDS) selects actions using both a measure of expected regret as well as a measure of expected information gain derived from a Bayesian model \citep{russo2014learningids}. 
 
Using the notation from \citet{russo2014learningids}, the action $A_{t}$ is a random variable depending on the history of observations 
$$\mathcal{F}_{t} = \sigma (A_{1}, r_{1,A_{1}}, \dots, A_{t-1}, r_{t-1, A_{t-1}} ).$$ 
Define a discrete distribution $\alpha_{t}$ on the space of actions $\mathcal A$, by $\alpha_{t} (a) =  \mathbb{P} \big(A^{*} = a \mid \mathcal{F}_{t} \big)$, where $A^{*}$ denotes the random variable taking on the value of the optimal action. The \emph{expected regret} obtained from taking action $a$ at step $t$ is denoted by $\nabla_{t}(a) = \mathbb{E} \big[ r_{t, A^{*}} - r_{t, a} \big]$. The \emph{information gain} obtained from taking action $a$ at step $t$ is denoted by $g_{t} (a)$ and is defined as the expected decrease in entropy of $\alpha_{t}$,  
\begin{equation*}
g_{t}(a) := \mathbb{E} \big[ H(\alpha_{t}) - H(\alpha_{t+1}) \mid \mathcal{F}_{t},  A_{t} = a \big].
\end{equation*}

Let $\mathcal{D} \big( \mathcal{A} \big)$ denote the set of distributions over the actions.  For a fixed $p \in \mathcal D( \mathcal{A})$ let $\nabla_{t}(p) := \sum_{a \in \mathcal{A}} p(a) \nabla_{t}(a)$ denote the expected regret when actions are selected by drawing from $p$. Similarly let $g_{t}(p) := \sum_{a \in \mathcal{A}} p(a) g_{t}(a)$ denote the expected information. The IDS policy samples actions from the distribution that minimises the ratio of the squared expected regret to the expected information gain, 
\begin{equation}
\label{ids-eq}
p_{t}^{\mathrm{IDS}} \in \underset{ p \in \mathcal{D} \big( \mathcal{A} \big) }{\operatorname{arg~min}}  \Bigg\{ \Phi_{t} (p) := \frac{\nabla_{t}(p)^{2}}{g_{t}(p)} \Bigg\}.
\end{equation}

It is known \citep{russo2014learningids} that this optimum is achieved at an extreme point of the simplex; furthermore, $g_{t}(a)$ is the mutual information between the two random variables $A^{*}$ and $r_{t, a}$ in the posterior distribution,
\begin{equation}
\label{kl-equiv}
\begin{aligned}
g_{t}(a) &=  I_{t} \big( A^{*}; r_{t,a} \big)   
= \mathbb{KL} \Big( \mathbb{P} \big( \big( A^{*}, r_{t,a} \big) \in \cdot \mid \mathcal{F}_{t} \big) \,\|\, 
\mathbb{P} \big(A^{*} \in \cdot \mid \mathcal{F}_{t} \big) \mathbb{P} \big( r_{t,a} \in \cdot \mid \mathcal{F}_{t} \big) \Big).
\end{aligned}
\end{equation}
This makes it possible to estimate the information gain by simulation. \citet{russo2014learningids} provide regret bounds for several general cases and gives examples of distributions for which the IDS policy is clearly superior to TS.

\begin{algorithm} [t]
\caption{Stochastic Variational Inference for Gamma Process Factor Analysis \label{svi algorithm}}
\begin{algorithmic}
\REPEAT
\FOR{ parameter $\theta$ in $\Theta=\{(W_{dk}),(r_{k}), \gamma, \gamma_{0}, (c_{0})\}$}
\STATE sample $z_\theta \sim N(0,1)$
\STATE set $\theta = \exp(\mu_\theta + z_\theta \sigma_\theta)$
\ENDFOR
\STATE
Formulate the ELBO gradient as an expectation over $(z_\theta)_{\theta\in\Theta}$, and compute estimate $\nabla \text{ELBO}$ by Monte Carlo with sample. 
\STATE
Compute step size $\Delta$ using AdaDelta.
\FOR{ parameter $\theta$ in $\Theta=\{(W_{dk}),(r_{k}), \gamma, \gamma_{0}, (c_{0})\}$}
\STATE
Update $\mu_{\theta} \leftarrow \mu_{\theta}+\Delta_{\mu_{\theta}}\nabla_{\mu_\theta}\text{ELBO}$
\STATE
Update $\sigma_{\theta} \leftarrow \sigma_{\theta}+\Delta_{\sigma_{\theta}}\nabla_{\sigma_\theta}\text{ELBO}$
\ENDFOR
\UNTIL{convergence}
\end{algorithmic}
\end{algorithm}

\section{Policies for Sequential Matrix Completion} 

In this section we describe an implementation of TS and IDS for sequential matrix completion. The prior on the matrix $M$ is a gamma process factor model, described by \citet{Knowles}. Efficient inference methods are critical, as these Bayesian policies require updating the posterior after every step. We employ a Stochastic Variational Inference algorithm.

The prior assumes that the columns of the user preference matrix $M$, denoted $m_1,\dots,m_N\in \mathbb R^d$, are drawn from a normal distribution $m_{n} \mid x_{n} \sim N(W x_{n}, \sigma^{2}I)$, where $x_{n} \sim N(0,I_{K})$. Integrating out $x_{n}$, this can be written $m_{n} \sim N(0, WW^{\top} + \sigma^{2} I)$. The prior on $W$ is specified by,
\begin{equation}
\begin{aligned}
W_{dk}  \mid r_{k}, \gamma  &\sim \textrm{G}(\gamma r_{k}, \gamma)  &  \quad
r_{k} \mid \gamma_{0}, c_{0} &\sim \textrm{G}(\gamma_{0}/K, c_{0}) & \\
\gamma &\sim \textrm{G}(1,1) & \quad
\gamma_{0} &\sim \textrm{G}(1,1) & \quad
c_{0} &\sim \textrm{G}(1,1),
\end{aligned}
\end{equation}
where G$(a,b)$ denotes the gamma distribution with density
$G(x; a,b)\propto x^{a-1} e^{-xb}$.

This model in practice assumes a large fixed value of $K$. However, the gamma process prior on $(r_k)$ tends to shrink the effective rank of the matrix $W$, and if the true number of factors $K^*$ is smaller than $K$, the posterior of the effective rank of $W$ concentrates around $K^*$. 

Recall that the observation at time $t$ is
\begin{equation}
Y_{t}=\text{tr}( Z_{t}^{\top}M )\beta^{T_{t,Z_t}}+\varepsilon_{t},
 \end{equation}
 where, again, $\textrm{tr}( Z_{t}^{\top}M )$ corresponds to some entry $m_{ij}$ of $M$. 
Since $\beta=0$, each entry is only observed once in a small-enough horizon. Because of this, we can take each $m_{ij}$ to be the noisy observation itself, absorbing the error variable $\varepsilon_t$ into the prior for $M$.

To choose an entry sequentially in TS or in IDS, we must sample the posterior distribution of the unobserved entries. 
Let $m_{nO}$ be the entries observed in column $n$, and $m_{nU}$ the rest of the entries in this column. As the columns of $M$ are conditionally independent given the parameters $W$ and $\sigma$,
 \begin{align}
&p(m_{1U},\dots,m_{NU}, W, \sigma \mid m_{1O}, \dots, m_{NO}) = \label{posterior} 
p(W, \sigma \mid m_{1O}, \dots, m_{NO})\prod_{n=1}^N p(m_{nU} \mid m_{nO}, W, \sigma).
\end{align}

Each factor $p(m_{nU} \mid m_{nO}, W, \sigma)$ on the right hand side is a normal distribution. Let 
\begin{equation}
\Sigma_{n}= \begin{bmatrix}
    \Sigma_{nUU}       & \Sigma_{nUO} \\
    \Sigma_{nOU}    &  \Sigma_{nOO} \\
\end{bmatrix}.
\end{equation}
be the covariance matrix $WW^\top+\sigma^2 I$ with rows and columns permuted such that the unobserved entries in $m_n$ appear first. It can easily be shown that 
\begin{equation}
\label{mvn posterior}
m_{nU} \mid m_{nO},W,\sigma \sim N(\bar \mu_{n}, \bar \Sigma_{n}),
\end{equation}
where 
\begin{equation}
\bar \mu_{n}= \Sigma_{nUO} \Sigma_{nOO}^{-1} m_{nO},
\quad \bar \Sigma_{n}= \Sigma_{nUU} + \Sigma_{nUO}\Sigma_{nOO}^{-1} \Sigma_{nOU}.
\end{equation}
This we sample efficiently using the trick described in \citet{doucet2010note}. The more difficult task is sampling the posterior of $W$ and $\sigma$ in the first factor of \ref{posterior}, which we approximate variationally.

More specifically, we compute a fixed-form, mean field variational posterior
\begin{equation}
\begin{aligned}
q^*(W,r,\gamma,\gamma_0,c_0) 
&= {\arg \min}_{q\in\mathcal Q} \;\text{KL}( p(W,r,\gamma,\gamma_0,c_0) \;\|\; q(W,r,\gamma,\gamma_0,c_0) ) \\
&= {\arg \max}_{q\in\mathcal Q} \; \mathbb{E}_q\left[ \log \frac{p(W,r,\gamma,\gamma_0,c_0)}{q(W,r,\gamma,\gamma_0,c_0)} \right] \label{ELBO}
\end{aligned}
\end{equation}
where $\mathcal Q$ is a parametric family of distributions in which the parameters $W_{dk}$, $r_k$, $\gamma$, $\gamma_0$, and $c_0$ for $d=1,\dots,D$ and $k=1,\dots,K$ are independent, and 
\begin{equation}
\begin{aligned}
\log W_{dk} \sim N(\mu_{W_{dk}}, \sigma^{2}_{W_{dk}}), & \qquad
\log r_{k} \sim N(\mu_{r_{k}}, \sigma^{2}_{r_{k}}), & \quad
\log  \gamma \sim N(\mu_{\gamma}, \sigma^{2}_{\gamma}), \\
\log  \gamma_{0} \sim N(\mu_{\gamma_{0}}, \sigma^{2}_{\gamma_{0}}), & \qquad
\log  c_{0} \sim N(\mu_{c_0}, \sigma^{2}_{c_{0}}).
\end{aligned}
\end{equation}

The maximisation objective in Eq. \ref{ELBO} is known as the evidence lower bound (ELBO) as it bounds the marginal probability of the data below. We rely on Stochastic Variational Inference \citep{SVI,SVIReview} to solve this problem. This method applies stochastic approximation algorithms to optimise the ELBO, deriving unbiased estimates of its gradient via Monte Carlo integration. More specifically, we apply the reparametrization trick introduced by \citet{salimans2013fixed} and \citet{kingma2013auto} to estimate the ELBO gradient, using the natural transformation mapping a standard normal to a log-normal. 
The choice of step size for the variational parameter updates is critical to the runtime of the algorithm. We use the AdaDelta method \citep{zeiler2012adadelta} to ensure fast convergence that is mostly unaffected by the initial choice of parameters. The whole procedure is summarized in Algorithm \ref{svi algorithm}.

We find improved performance with a good initialisation. This is achieved by maximising the log posterior of the parameters $W_{dk}, r_{k}, \gamma, \gamma_{0}, c_{0}$ and centering the initial variational posterior around the maximum a posteriori estimate.

TS only requires samples from the posterior. On the other hand, IDS requires estimating the information ratio in Eq.\ \ref{ids-eq}. Algorithm $\ref{approxIR}$, drawn from \citet{russo2014learningids}, approximates the information-ratio in Eq. $\ref{ids-eq}$. This algorithm uses the equivalence presented in Eq. $\ref{kl-equiv}$ in that it consists of approximating the $\mathbb{KL}$ divergence of the distributions $\mathbb{P} \big( \big( A^{*}, r_{t,a} \big) \in \cdot \mid \mathcal{F}_{t} \big)$ and  $ \mathbb{P} \big(A^{*} \in \cdot \mid \mathcal{F}_{t} \big) \mathbb{P} \big( r_{t,a} \in \cdot \mid \mathcal{F}_{t} \big) $. Let $\theta$ denote the covariance matrix $WW^{\top}+\sigma^{2}I$ drawn from the posterior distribution $\pi(\theta)$. Let $f_{\theta, a} (y)$ be the probability of observing value $y$ when taking action $a$ conditioned on $\theta$. Let $p(a^{*})$ be a discrete approximation to $\alpha_{t} (a) =  \mathbb{P} \big(A^{*} = a \mid \mathcal{F}_{t} \big)$, $p_{a}(y)$ to the probability of observing $y$ from action $a$, and $p_{a}(a^{*}, y)$ to the probability of observing $y$ from action $a$ when action $a^{*}$ is optimal. Then one can check that 
\begin{equation*}
\begin{aligned}
 \mathbb{E} \big[ H(\alpha_{t}) - H(\alpha_{t+1}) \mid \mathcal{F}_{t},  A_{t} = a \big] 
&= \mathbb{KL} \Big( \mathbb{P} \big( \big( A^{*}, r_{t,a} \big) \in \cdot \mid \mathcal{F}_{t} \big) \,\|\, 
 \mathbb{P} \big(A^{*} \in \cdot \mid \mathcal{F}_{t} \big) \mathbb{P} \big( r_{t,a} \in \cdot \mid \mathcal{F}_{t} \big) \Big) \\
& \approx \sum_{a^{*}, y} p_{a}(a^{*}, y) \log  \frac{p_{a}(a^{*},y)}{p(a^{*})p_{a}(y)}.
\end{aligned}
\end{equation*}


\begin{algorithm}[t]
   \caption{
   Thompson Sampling}
   \label{alg:example}
\begin{algorithmic}
   \STATE {\bfseries Input:} N, P, some upper bound on the rank, K, some upper bound on the variance, $\sigma$, and the discount factor $\beta=0$.
   \STATE Sample small percent of $K^*(N+D-K^*)$ entries uniformly with replacement
   \REPEAT
   \STATE Update model parameters to maximise  $\log p(Y_{1:t},  W, r, \gamma, \gamma_{0}, c_{0}, \sigma) $
   \STATE Maximise ELBO through SVI
   \FOR{$i=1$ {\bfseries to} $N_{\textrm{obs}}$} 
   \STATE Draw $W,\sigma$ from variational approximation $q(W, \sigma \mid Y_{1:t})$
   \STATE For each $n$, draw $m_{n}$ from $p(m_{nU} \mid m_{nO}, W, \sigma) p(m_{nO}\mid W,\sigma, Y_{1:t})$
   \STATE Compute discounted reward for each entry, $m_{nk} \beta^{N_{m_{nk}}}$, where $N_{m_{nk}}$ is the number of times entry $(n,k)$ has been observed
   \STATE Observe the entry with largest reward
   \ENDFOR
   \UNTIL{Matrix can be completed via convex optimisation}
\end{algorithmic}
\end{algorithm}

\begin{figure*}
\centering
\includegraphics[width=0.45\linewidth]{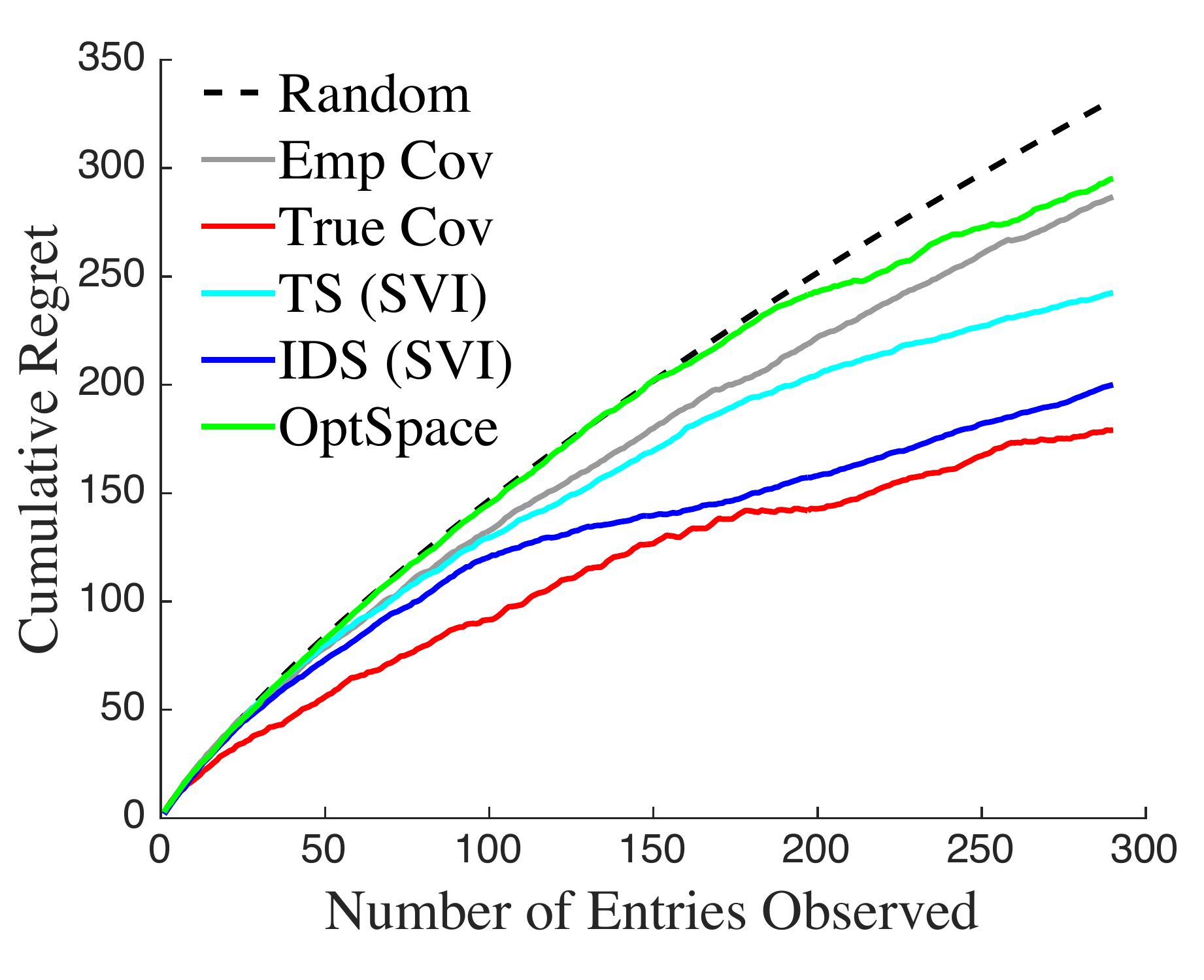}
\includegraphics[width=0.45\linewidth]{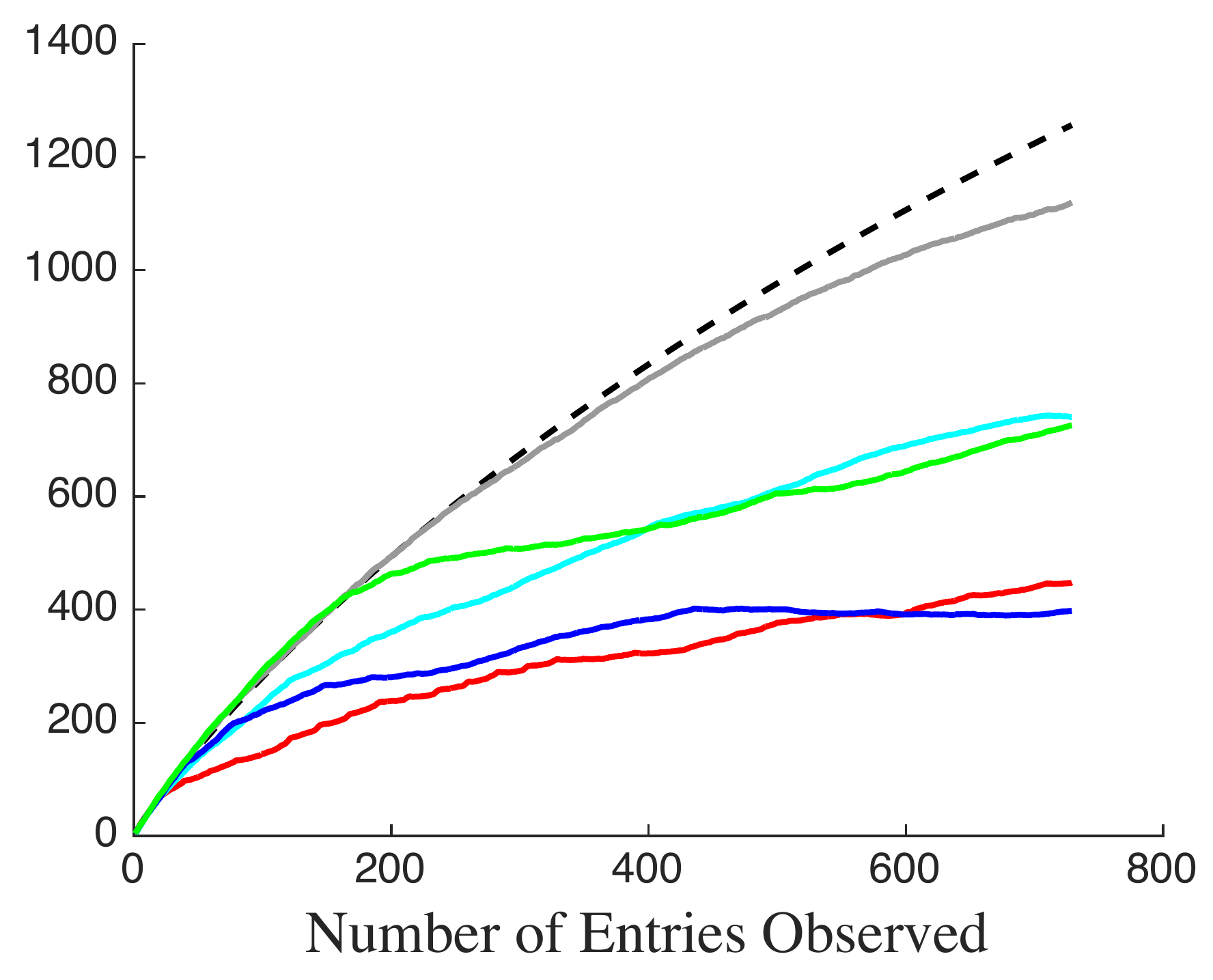}
\includegraphics[width=0.45\linewidth]{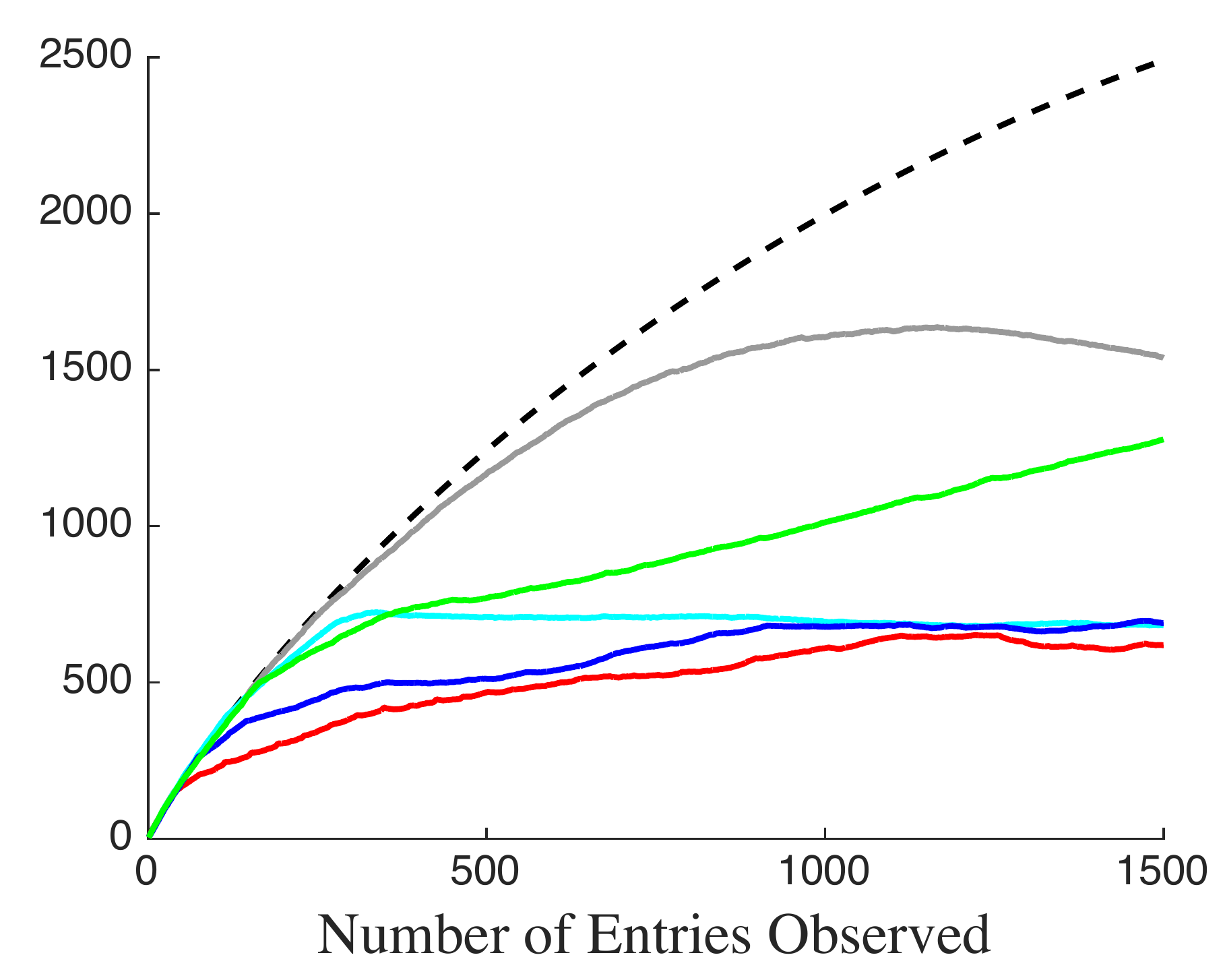}
\includegraphics[width=0.45\linewidth]{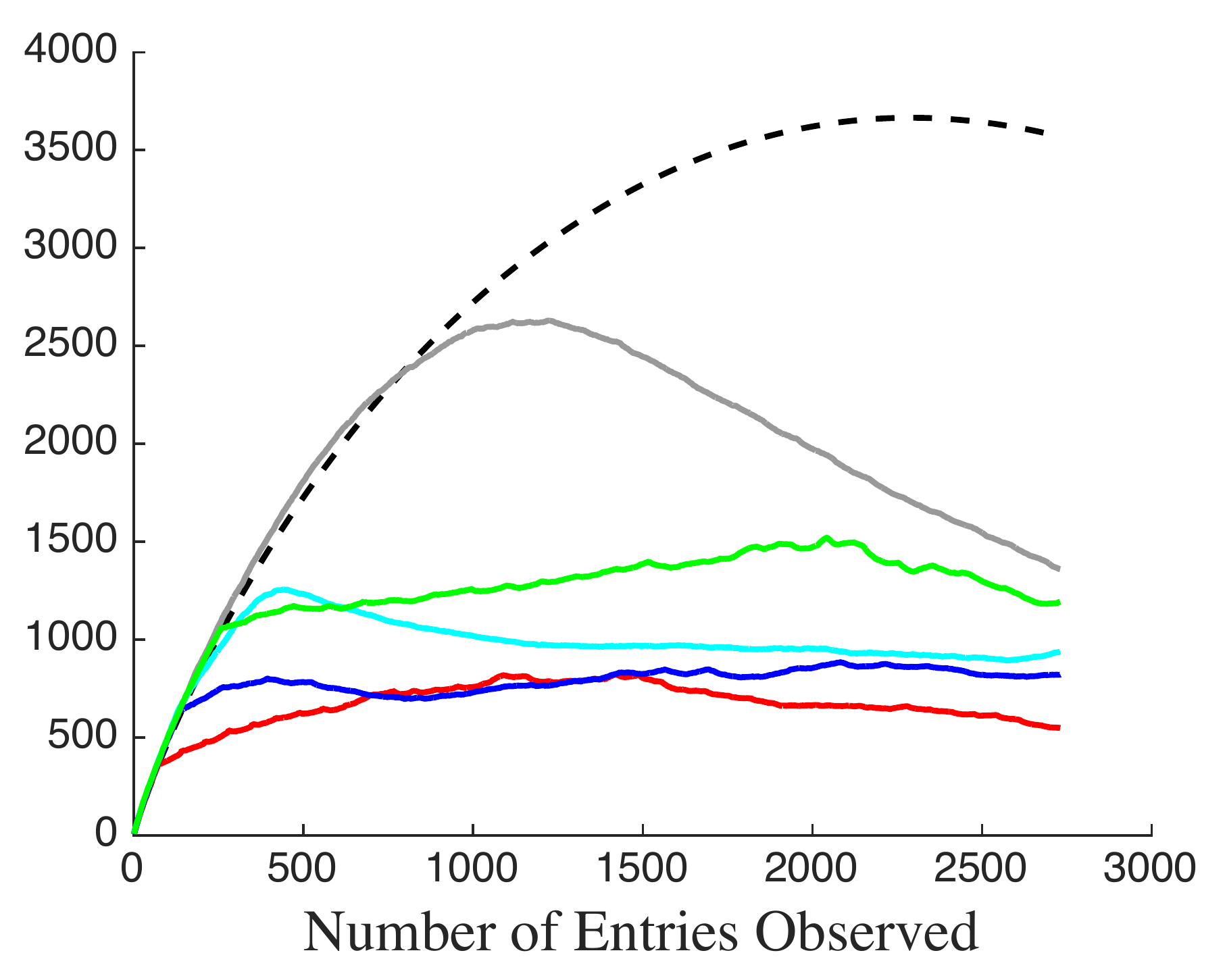}
\caption[width=0.3\linewidth]{Regret with synthetic data. The true rank of $W$ is 1 (top left), 5 (top right), 10 (bottom left), and 20 (bottom right). \label{synthetic data figure}}
\end{figure*}

\section{Results}

\subsection{Synthetic Data}
To test the performance of each algorithm we construct a $D \times K^*$ factor matrix $W$, for ranks $K^*=1,5,10,15,20$, $D=50$, and $\beta=0$. For half of the runs, the true factor matrix $W$ is sampled elementwise uniformly on $[0,1]$, otherwise it is sampled elementwise from $\textrm{Beta}(2,5)$.  We draw $N=100$ columns from a multivariate normal with covariance matrix $WW^{\top}+0.1 I$. We average the results over 10 runs and measure the regret by constructing the optimal sequence of actions knowing the full true matrix and subtracting the cumulative reward from the cumulative reward achieved by the policy of each algorithm. For these Bayesian methods we set the user-defined threshold rank to be $ K^{*}= 20$. Every test run considers the horizon $H=K^* (N+D- K^*)$, since after which many methods can complete the full matrix \citep{Candes-Noise}. For computational efficiency, rather than observing a single entry at a time and updating the posterior at each iteration, we observe $N_{\textrm{Obs}} = H/40$ entries between updates of the posterior.

The regret curves are shown in Fig.\ \ref{synthetic data figure}. The plots include the result of an oracle policy which uses the true covariance matrix $WW^\top$ to sample the multivariate normal posterior (\ref{mvn posterior}) of the missing entries, as well as a policy using an empirical estimate of the covariance matrix in a similar way, and a greedy policy relying on matrix completion via OptSpace \citep{keshavan2009matrix}. The IDS policy performs almost as well as the oracle policy, and TS is superior to the policies using the OptSpace estimate and the empirical covariance. 

\subsection{MovieLens and Jester Data}
\begin{figure}[t]
\includegraphics[width=0.5\linewidth]{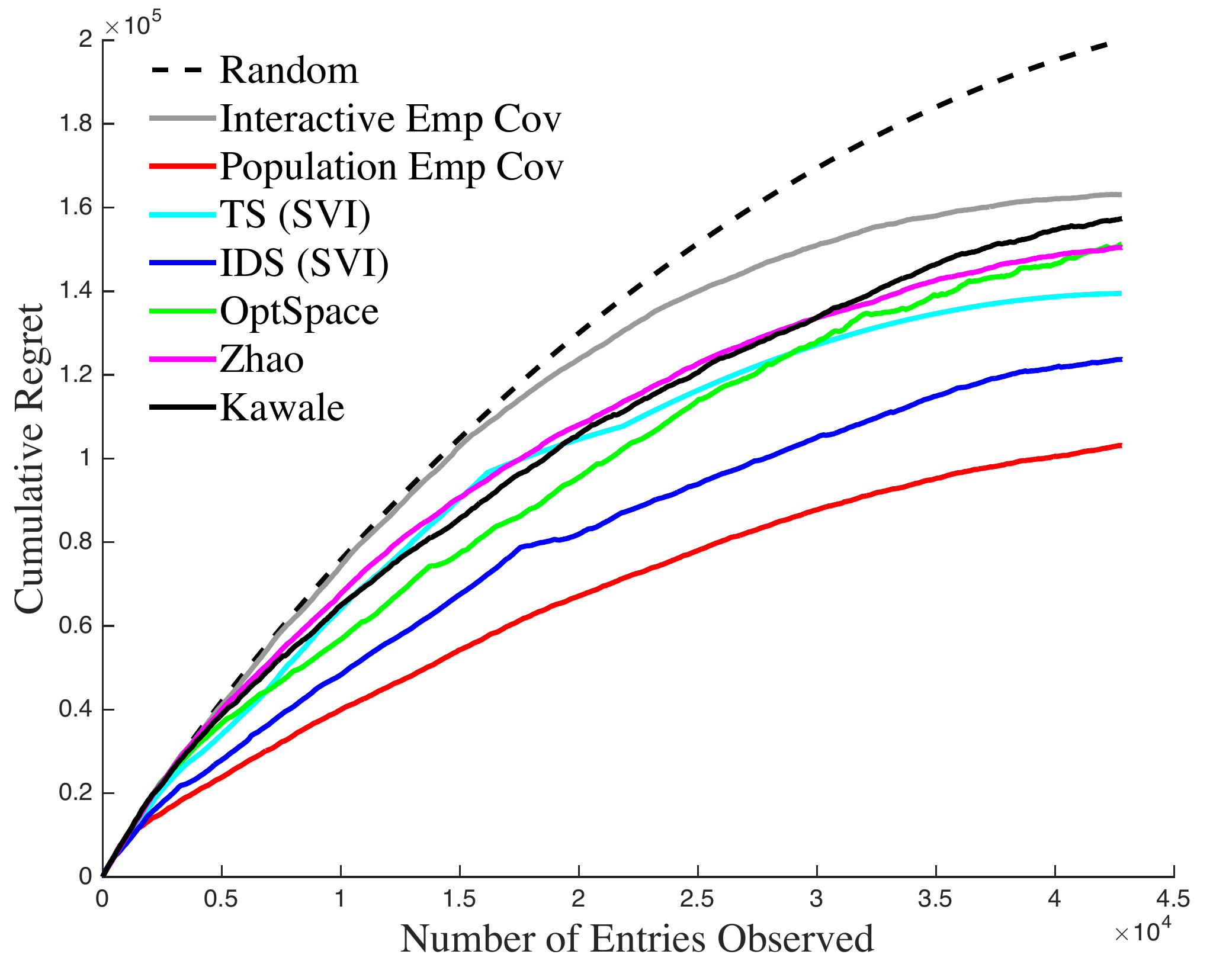}
\includegraphics[width=0.5\linewidth]{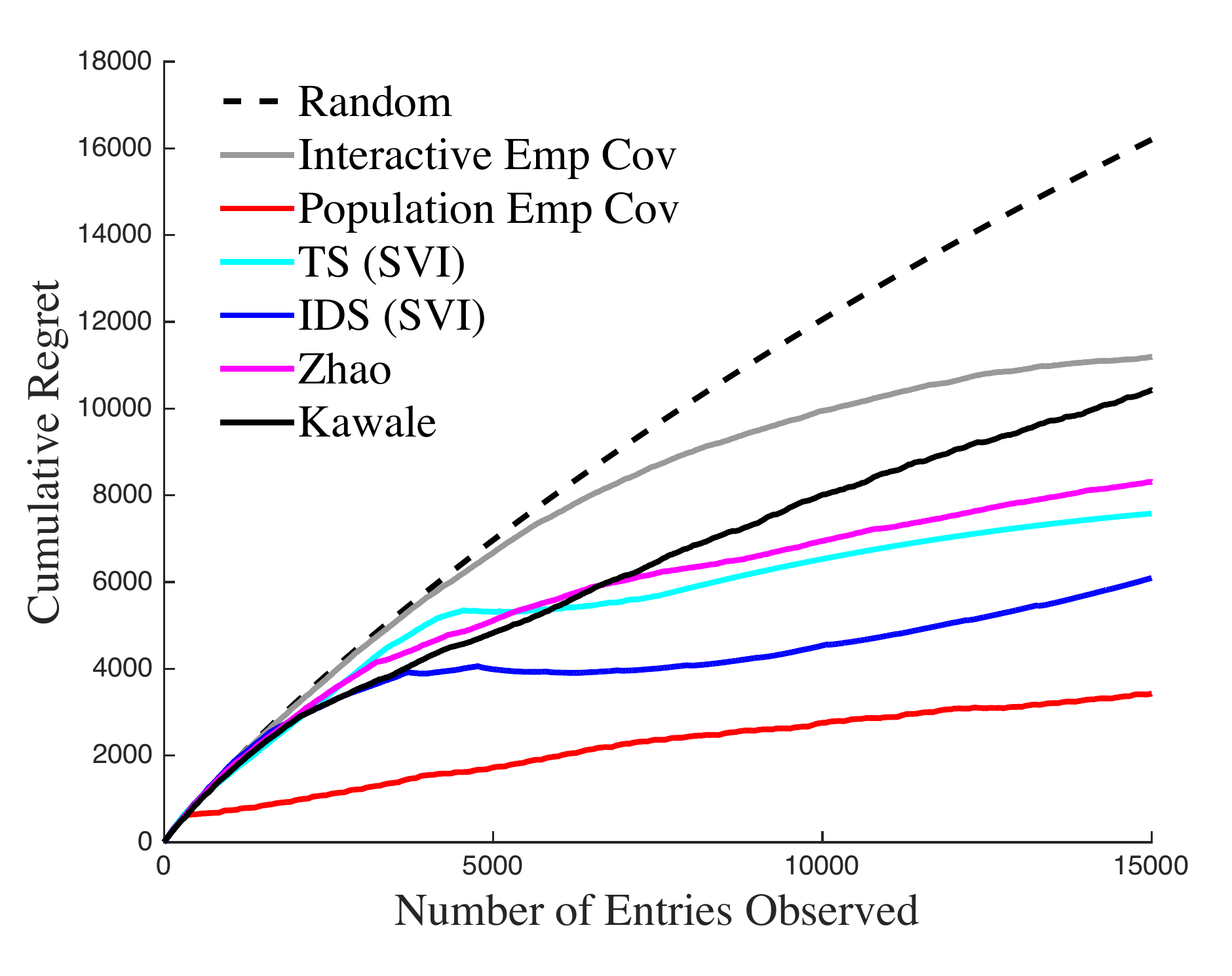}
\caption[width=0.3\linewidth]{Regret in experiment with Jester dataset (left) and MovieLens dataset (right). \label{real regret figure} }
\end{figure}

We test the algorithms on both a $1000 \times 100$ matrix from the MovieLens dataset \citep{1} and the Jester dataset \citep{goldberg2001eigentaste}. For the MovieLens dataset, in order to be able to compute the regret, we complete the missing entries with OptSpace \citep{keshavan2009matrix}, an algorithm known to perform well on this dataset \citep{keshavan2009low}, and treat the completed version as the true ratings matrix. The Jester dataset has no missing entries so we compare OptSpace's performance to our algorithms. We consider the horizon $K^*(N+D-K^*)$, with $N=1000$, $D=100$, and $ K^{*}=20$ for both datasets. For comparison we use a method that, at each step, updates the empirical covariance matrix and then draws estimates of the unobserved entries from the corresponding conditional multivariate normal pseudo-posterior. As a near-oracle comparison algorithm we first compute the `population' covariance matrix using the full dataset with no missing entries, then sequentially draw missing entries from the corresponding multivariate normal pseudo-posterior. Finally we compare the performance of these algorithms to the current competitive methods from  \citet{rctest} and \citep{zhao2013interactive}.The regret curves are shown in Fig.\ \ref{real regret figure}.

\begin{algorithm}[h]
   \caption{\texttt{ApproxInfoRatio($\pi$)} from \citep{russo2014learningids}}
   \label{approxIR}
\begin{algorithmic} 
   \STATE Draw samples $\Theta=(\theta_i)$ i.i.d. $\theta_i\sim \pi$.
   \STATE Construct histogram bins for values of $y$ to use henceforth.
      \STATE Compute a discrete approximate to $f_{\theta, a} (y)$.
         \STATE $\Theta_{a} \leftarrow \left \{ \theta \mid a = \textrm{arg max}_{a'} \sum_{y} q_{\theta, a'} (y) y \right \}$ 
   \STATE $p(a^{*}) \leftarrow \sum_{\theta \in \Theta_{a^{*}}} \pi(\theta)$ $\forall a$
      \STATE $p_{a}(y) \leftarrow \sum_{\theta} \pi(\theta) q_{\theta, a} (y)$ $\forall a, y, \theta$
   \STATE $p_{a}(a^{*}, y) \leftarrow \frac{1}{p(a^{*})} \sum_{\theta \in \Theta_{a^{*}}} q_{\theta, a}(y)$
   \STATE $R^{*} \leftarrow \sum_{a}\sum_{\theta \in \Theta_{a}} \sum_{y} \pi(\theta) q_{\theta, a} (y) R(y)$ where $R(y) = y \beta^{T_{t,a_{t}}}$ (i.e. incorporate discount factor into the rewards).  
   \STATE $g_{a} \leftarrow \sum_{a^{*}, y} p_{a}(a^{*}, y) \log  \frac{p_{a}(a^{*},y)}{p(a^{*})p_{a}(y)}$ $\forall a$
   \STATE $\nabla_{a} \leftarrow R^{*} - \sum_{\theta} \pi(\theta) \sum_{y} q_{\theta, a}(y) R(y)$
      \STATE Output matrices $\bar \nabla$, $\bar g$

\end{algorithmic}
\end{algorithm}

\begin{algorithm}[h]
   \caption{
   Information-Directed Sampling}
   \label{ids_alg}
\begin{algorithmic}
    \STATE {\bfseries Input:} N, P, some upper bound on the rank, K, some upper bound on the variance, $\sigma$, and the discount factor $\beta =0$.
   \STATE Sample small percent of $K^*(N+D-K^*)$ entries uniformly with replacement
   \REPEAT
   \STATE Update model parameters to maximise $\log p(Y_{1:t},  W, r, \gamma, \gamma_{0}, c_{0}, \sigma)$
   \STATE Maximise ELBO through SVI
   \STATE Compute $\bar \nabla, \bar g$ from \texttt{ApproxInfoRatio}. 
 \FOR{$i=1$ {\bfseries to} $N_{\textrm{obs}}$} 
   \STATE Select action $a \in \textrm{arg min} \,\bar \nabla^{2} ./ \bar g$ (where $./$ denotes elt-wise division) to observe.
   \STATE Set $\nabla_{a} \leftarrow R^{*} - \sum_{\theta} \pi(\theta) \sum_{y} q_{\theta, a}(y) y \beta^{T_{t,a_{t}}}$.
   \ENDFOR
   \UNTIL{Matrix can be completed via convex optimisation}
\end{algorithmic}
\end{algorithm}

\section{Discussion}

\subsection{Related Work}
Our work falls into the area of collaborative filtering. Recently, the idea of applying bandit algorithms to online collaborative filtering has become more prevalent \citep{li2010contextual,bresler2014latent,li2016collaborative,deshpande2012linear}. However, much of the literature focuses on contextual bandits, in which a user's preference is a function of a given set of predictors. The setting in which observations are corrupted entries of a low-rank matrix is less common and less amenable to theoretical analysis due to the fact that adaptive confidence intervals are more difficult to obtain. Applying TS to a collaborative filtering application was first proposed by \citep{zhao2013interactive}. The authors use the PMF model and employ Markov chain Monte Carlo and Gibbs Sampling to sample from the posterior distribution. In a similar vein, \citep{rctest} employs a PMF model and implements a ``Rao-Blackwellized particle filter'' to give a discrete approximation for the posterior distribution. Here we use a gamma process factor model which has the advantage of adapting to the true rank of the data. The main contribution of this paper is a demonstration that crude uncertainty estimates from SVI, coupled to smart Bayesian policies like IDS, can lead to near optimal designs for horizons smaller than the typical matrix completion threshold.  There are several avenues to extend the work presented here. Among them, we highlight investigating the performance of the policies under different models of corruption or constraints on the set of available actions, and the integration of other predictors into the low-rank model.

\subsection{Future Work}
\subsubsection{Computational Efficiency}
While SVI makes it possible to implement Bayesian policies in an online collaborative filter, the computational cost is still significant. IDS outperforms TS but involves a heavier computational burden. Applying these policies in real-world settings would require further research into fast inference algorithms or simplified uncertainty estimates. In particular, developing models in which the dimension of the parameter grows sublinearly in $N$ and $D$ is critical to make our policies feasible in a big-data application. In the Bayesian framework for sequential matrix completion there is a trade-off between computation time and performance. Information Directed Sampling outperforms TS, however it is much slower to compute and much less scalable. Given the impressive performance of IDS in both the synthetic data and real data, it seems that future research into how to improve the computational complexity of the method would be worthwhile. Perhaps there is a suitable, more computable alternative that is also able to quickly identify which actions lead to the best information about the underlying Bayesian model.

\subsubsection{Regret Bounds}
Deriving regret bounds for our policies would require significant technical advances. The methods in \citet{russo2014learning,russo2014learningids} can be used to obtain weak bounds on the Bayes regret, but sharp bounds which scale optimally in the number of degrees of freedom of the true matrix  would require efficient confidence intervals for the missing entries in the matrix. The main hindrance is that most of the theory of low-rank matrix completion relies substantially on random designs. The bounds in \citet{klopp2014noisy} for non-uniform sampling could potentially be useful even though they still require independence in the design. A line of recent work on matrix completion with deterministic sampling could also provide tools for regret bounds \citep{kiraly2015algebraic, pimentel2016characterization}; however, there is work to be done to sharpen these results and translate computable certificates of completability into simple conditions. In a previous analysis of TS for sequential matrix completion, \citet{rctest} asserted that new tools to analyze generic posterior distributions are needed for robust regret bounds. However they show how to bound the regret in the special case of rank-1 matrices. 

Our model induces a duality between estimating the missing entries and estimating the covariance of the columns. Unfortunately, much of the theory of covariance estimation in high-dimensional statistics is limited to the case in which i.i.d. vectors without any missing entries are observed. An exception to this can be seen in \citet{lounici2014high}; however this analysis still relies on sampling entries uniformly at random. Guarantees on covariance estimation given structured sequences of partial observations would be essential to deriving regret bounds.

\end{document}